\documentclass[proceedings]{JHEP} % 10pt is ignored!

\usepackage{epsfig}
\usepackage{latexsym}
\usepackage{amsfonts}
\usepackage{amssymb}			
\newbox\mybox
\newcommand{\ttbs}{\char'134}           % \backslash for \tt (Nucl.Phys. :)%
\newcommand\fverb{\setbox\mybox=\hbox\bgroup\verb}
\newcommand\fverbdo{\egroup\medskip\noindent\fbox{\unhbox\mybox}\ }
\newcommand\fverbit{\egroup\item[\fbox{\unhbox\mybox}]}

\font\beeg=cmr17 scaled 1600		% Stylish initials
\newcommand\init[1]{\setbox\mybox=\hbox{{\beeg #1}~}%
		   \noindent\global\hangindent=\wd\mybox\global\hangafter-2%
		   \sc\smash{\llap {\lower 13.2pt \box\mybox}}}
\title{Superspace Representations of SU(2,2/N) Superalgebras and Multiplet
Shortening}

\author{S. Ferrara%\thanks{A footnote may follow the name of the speaker
		\\
	Theoretical Physics Division, CERN, Geneva 23, Switzerland\\
	E-mail: \email{Sergio.Ferrara@cern.ch}}

\conference{Quantum Aspects of Gauge Theories, Supersymmetry and Unification}

\abstract{We derive massless and massive representations of all SU(2,2/N)
superalgebras by using superfields defined in ``harmonic superspace".  This
method allows one to easily construct ``short superfields" which are relevant
in the analysis of the AdS/CFT correspondence.}

\begin{document} 

\maketitle %%%%%%%%%% THIS IS IGNORED %%%%%%%%%%%
%%%%%%%%%%%%%%%%%%%%%%%%%%%%%%%%%%%%%%%%%%%%%%%%%%%%%%%%%%%%%%%%%%%%%%%%%%%%%
\section{Introduction}

The study of superconformal algebras has recently attracted 
renewed interest for their dual role in the ${\rm AdS}_{d+1}/{\rm 
CFT}_d$ correspondence \cite{mal,gkp,wit}, connected to the 
near-horizon geometry of $d-1$-branes. 

A special role is played by $3$-branes since they are related to 
superconformal invariant quantum Yang-Mills theories. These 
theories are the only ones exhibiting conformal symmetry both at 
weak and strong coupling and, in any case, admitting, unlike other 
types of branes, Yang-Mills fields in the conformal regime. 

The bulk and boundary operators in this correspondence are 
classified by highest weight UIR's of $SU(2,2/N)$ algebras 
\cite{fefrza,AF} where $N=1,2$ and 4 in the known examples, since 
supergravity or superstring theory can admit at most 32 ($8N$) 
supersymmetries. Nevertheless, in the study of superconformal 
algebras and their representations different values of $N$ are of 
interest because they help one to exhibit some general features of 
short representations, corresponding to conformal operators with 
protected dimension, but more importantly, because these algebras 
may be relevant for some generalizations of the known schemes in 
which more than 32 supersymmetries may be required \cite{bgg1}. 

Recently \cite{AFSZ} it has been shown that a known 
generalization of ordinary superspace, called ``harmonic 
superspace" \cite{GIK1}-\cite{GIK3}, is particularly suitable to 
build up, in a rather simple and general manner, all possible 
composite operators of superconformal invariant gauge theories 
with $N>1$ extended supersymmetry. 

Other approaches, like ordinary superspace \cite{Siegel,HST} or 
the oscillator construction \cite{bgg}-\cite{gmz1} of highest 
weight representations, although in principle possible, are much 
more complicated to deal with and the complete analysis of all 
possible shortenings would be unnecessarily difficult. 

In fact, the structure of harmonic superspace is powerful enough 
to allow us to extend the analysis of Ref. \cite{AFSZ} to all 
$SU(2,2/N)$ superalgebras  with arbitrary $N$,\footnote{For a 
thorough treatment of the action of superconformal groups on 
harmonic superspaces see Refs. \cite{hh}.} although no dynamical 
theory is known for $N>4$. This report contains results obtained in 
Ref.~\cite{sf}.

{}From a mathematical point of view harmonic superspace is an 
enlarged space where superfields are defined on ``flag manifolds" 
\cite{hh,knapp} 
\begin{equation}\label{flag}
  {\mathcal M} = \frac{SU(N)}{S\left(U(n_1)\times\ldots\times
U(n_p)\right)}\;, 
 \left(\sum_{k=1}^p n_k =N\right). 
\end{equation}
To study the general case of multiplets it is important 
\cite{AFSZ} to use the choice $n_k=1$ ($k=1\ldots N$), i.e. where 
we quotient the group $SU(N)$ by its maximal torus. Then the above 
manifold is the largest flag manifold with complex dimension \\ \hfill 
$N(N-~1)/2$. 

The ultrashort UIR's of $SU(2,2/N)$ superalgebras described by 
analytic harmonic superfields depend only on half of the odd 
coordinates (Grassmann or G-analyticity):
\begin{equation}\label{-5}
   W^{12\ldots k} =  W^{12\ldots k}(\theta_{k+1},\theta_{k+2},\ldots,\theta_{N},
\bar\theta^1,\bar\theta^2,\ldots,\bar\theta^k)\;. 
\end{equation}
In addition, they are annihilated by all the ``step-up" generators 
$E_a$ in the Cartan decomposition of the Lie algebra of $SU(N)$. 
In other words, these superfields correspond to highest weight 
states of $SU(N)$: 
\begin{equation}\label{HW}
  E_a|{\rm HW}\rangle = 0\;.
\end{equation}
In harmonic superspace this irreducibility condition corresponds 
to harmonic (or H-) analyticity.  The crucial point is that the 
$SU(2,2/N)$ algebra acting on such states defines a 
``quasi-primary" superconformal field denoted by 
\begin{equation}\label{calD}
{\cal D} (\ell,J_1,J_2;r;a_1,\ldots, a_{N-1}) 
\end{equation}
where $\ell,J_1,J_2$ are the conformal dimension and spin of the 
state, $r$ is the $U(1)$ $R$ charge and $a_1,\ldots, a_{N-1}$ are 
the $SU(N)$ Dynkin labels. We assign the $R$ charge 
$r_\theta={1\over 2}(1-{4\over N})$ to the Grassmann coordinates 
in order to be consistent with the convention that chiral 
superfields $\Phi(\theta)$ have $l=-r$ for any $N$. This is also 
the charge which naturally appears in the definition of the   \\ 
$SU(2,2/N)$ superalgebra \cite{FKTN}. \footnote{Note that for 
$N=4$, $r_\theta =0$ and the $r$ quantum number becomes a 
``central charge" \cite{dp,bin}. In this case the analysis of 
section 2 refers to the $PSU(2,2/4)$ algebra for $r=0$ and to the 
$PU(2,2/4)$ algebra for $r\neq 0$.} 

The G- and H-analytic superfields (\ref{-5}) have their lowest 
(scalar) component belonging to the rank $k$ antisymmetric 
representation of $SU(N)$ ($k=1\ldots[{N\over 2}]$), have $R$ 
charge $r_k={2k\over N}-1$ and will be shown to describe 
``ultrashort" representations of the $SU(2,2/N)$ superalgebra. If 
the algebra is interpreted as acting on ${\rm AdS}_5$, these 
are the ``supersingleton" representations \cite{ff2}. For $k=0$ the 
superfield is actually ``chiral" and in this case the highest 
weight state may carry a spin label $(J_L,0)$ with $\ell =1+J_L$. 
The chiral superfield is the supersingleton representation when 
the top spin is $J_L={N\over 2}$. For all other analytic 
superfields $(k>1)$ the supersingleton will have top spin 
$J_L={N\over 2}-{k\over 2}$. 

It should be pointed out that the same massless multiplets can be 
described in terms of ordinary but {\it constrained} superfields 
\cite{Siegel, HST}. The reason why we prefer the harmonic 
superspace version is the fact that the superfields  (\ref{-5}) 
are {\it unconstrained} analytic objects. Analyticity is a 
property which is preserved by multiplication. This will allow us 
to tensor the above massless UIR's in a very simple way and thus 
obtain series of short multiplets of $SU(2,2/N)$. We observe that 
from the ${\rm AdS}_5$ point of view, tensoring more than two 
supersingleton reps produces ``massive bulk"  reps, while 
tensoring only two of them produces ``massless bulk"  reps 
\cite{ff2,gmz1}. The latter are the ``supercurrent" multiplets 
discussed in Ref. \cite{HST}.

\section{Unitarity bounds and shortening of UIR's of $SU(2,2/N)$} 

The unitarity bounds of highest weight UIR's of $SU(2,2/N)$ have 
been derived in Refs. \cite{ff,dp,bin,mss}. They correspond to some bounds on 
the highest weight state (\ref{calD}). Let us define the 
quantities
\begin{equation}\label{mm}
  m_1=\sum_{k=1}^{N-1}a_k\;, \quad m=\sum_{k=1}^{N-1}(N-k)a_k
\end{equation}
and
\begin{eqnarray}\label{XY}
  X(J,r,{2m\over N}) &=& 2+2J-r+{2m\over N}\;, \nonumber \\ 
  Y(r,{2m\over N}) &=& -r+{2m\over N}\;.
\end{eqnarray}
Then we have $(J_1=J_L, \ J_2=J_R)$:
\begin{equation}\label{A}
  {\rm A)}\qquad \ell \geq X(J_2,r,{2m\over N}) \geq X(J_1,-r, 2m_1-{2m\over 
N})
\end{equation}
(or $J_1\rightarrow J_2$, $r\rightarrow -r$, ${2m\over 
N}\rightarrow 2m_1-{2m\over N}$);
\begin{equation}\label{B}
  {\rm B)}\qquad \ell = Y(r,{2m\over N}) \geq X(J_1,-r, 2m_1-{2m\over 
N}) 
\end{equation}
(or $J_1\rightarrow J_2$, $r\rightarrow -r$, ${2m\over 
N}\rightarrow 2m_1-{2m\over N}$); 
\begin{equation}\label{C}
  {\rm C)}\qquad \ell = m_1\;, \quad r={2m\over N}-m_1\;, \quad J_1=J_2=0\;. 
\end{equation}

The massless UIR's correspond to B) for $a_k=0$, $\ell=-r=1+J_L$ 
and to C) for $\ell=m_1=1$, $r_k= {2k\over N}-1$, $1\leq k \leq 
[{N\over 2}]$. Note that the two series overlap for $J_L=0$ in B) 
and $k=0$ in C).

The short multiplets that we shall build in section 4 by tensoring 
massless multiplets from the C) series in the case of $N=2n$ for  
$k=n$ ($r=0$)  will belong to the shortenings in B) and C) 
obtained for $J_1=0$ and $r=0$: 
\begin{eqnarray}
  &{\rm B)}&\ \ell= {2m\over N}\;, \quad  {2m\over N}-m_1 \geq 1\;\nonumber\\
  &{\rm C)}&\ \ell= m_1\;, \quad  {2m\over N}=m_1 \label{BC}
\end{eqnarray}
 
\section{Massless superconformal multiplets}

\subsection{Grassmann analytic superfields}

We consider superfields $$W^{i_1\ldots i_k}(x_{\alpha\dot\alpha}, 
\theta^\alpha_i,\bar\theta^{\dot\alpha i})$$ with $k=1,\ldots, n$ 
(where $n=[{N\over 2}]$) totally antisymmetrized indices in the 
fundamental representation of $SU(N)$. These superfields satisfy 
the following constraints:
\begin{eqnarray}
  &&D^{(j}_\alpha W^{i_1)i_2\ldots i_k}=0\;, \label{1-1}\\
  &&\bar D_{\dot\alpha \{j}W^{i_1\}i_2\ldots i_k}=0  \label{1-2} 
\end{eqnarray}
where $()$ means symmetrization and $\{\}$ means the traceless 
part. The spinor derivatives algebra is 
\begin{equation}\label{spder}
  \{D^{i}_\alpha,\bar D_{\dot\alpha j}\} = i\delta^i_j 
\partial_{\alpha\dot\alpha}
\end{equation}
with $\partial_{\alpha\dot\alpha} = 
\sigma^\mu_{\alpha\dot\alpha}\partial_\mu$. In the cases $N=2,3,4$  
these constraints define the on-shell $N=2$ matter 
(hyper)multiplet \cite{Sohnius} and the $N=3,4$ on-shell 
super-Yang-Mills multiplets \cite{So}. Their generalization to 
arbitrary $N$ has been given in Refs. \cite{Siegel,HST} where it 
has also been shown that they describe on-shell massless 
multiplets.  

Our aim in this section is to rewrite the constraints (\ref{1-1}), 
(\ref{1-2}) in harmonic superspace where they will take the simple 
form of analyticity conditions. Using this fact we will then be 
able to construct tensor products of the corresponding multiplets 
in a very straightforward and easy way (section 4). 

The main purpose of introducing harmonics is to be able to 
covariantly project all the $SU(N)$ indices in (\ref{1-1}), 
(\ref{1-2}) onto a set of $U(1)$ charges. To this end we choose 
the harmonic coset $SU(N)/(U(1))^{N-1}$ described in terms of 
harmonic variables $u^I_i$ and their conjugates $u^i_I = (u^I_i)^* 
$. \footnote{The harmonic notation used here differs from the 
original one of Refs. \cite{GIK1,GIK2}. It is similar to the one 
introduced in Ref. \cite{GIK3} for the case $N=3$ and in Refs. 
\cite{hh} for general $N$.} They form an $SU(N)$ matrix where $i$ 
is an index in the fundamental representation of $SU(N)$ and 
$I=1,\ldots,N$ is a collection of the $N-1$ $U(1)$ charges 
corresponding to the projections of the second index (the harmonic 
$u^i_I$ carries charges opposite to those of $u^I_i$). They 
satisfy the following $SU(N)$ defining conditions: 
\begin{eqnarray}
 &&u^I_i u^i_J=\delta^I_J~, \label{a2}\\ u\in 
SU(N):\quad &&u^I_i u^j_I =\delta^j_i~,\label{a2'}\\ && 
\varepsilon^{i_1\ldots i_N}u^1_{i_1}\ldots u^N_{i_N}=1~. 
\label{a2''} 
\end{eqnarray}

Now, let us use these harmonic variables to split all the $SU(N)$ 
indices in the constraints (\ref{1-1}), (\ref{1-2}) into 
independent $(U(1))^{N-1}$ projections. For example, the 
projection 
\begin{equation}\label{2}
  W^{12\ldots k} = W^{i_1i_2\ldots i_k} u^1_{i_1}u^2_{i_2}\ldots u^k_{i_k}
\end{equation}
satisfies the constraints
\begin{eqnarray}
&&D^{1}_\alpha  W^{12\ldots k} = 
D^{2}_\alpha  W^{12\ldots k} =\ldots =  \nonumber \\
&&D^{k}_\alpha  W^{12\ldots k} = 0\;, 
 \nonumber \\
\label{4a}
\end{eqnarray}
\begin{eqnarray}
&&  \bar D_{\dot\alpha\; k+1} W^{12\ldots k} = 
 \bar D_{\dot\alpha\; k+2} W^{12\ldots k} =\ldots =  \nonumber \\
&&\bar D_{\dot\alpha\; N} W^{12\ldots k} = 0 
\label{4b}
\end{eqnarray}
where $D^I_\alpha = D^i_\alpha u_i^I$ and $\bar D_{\dot\alpha\; 
I}= \bar D_{\dot\alpha\; i}u^i_I$. The first of them, eq. 
(\ref{4a}), is a corollary of the commuting nature of the 
harmonics variables, and the second one, eq. (\ref{4}), of the 
unitarity condition (\ref{a2}). The main achievement in rewriting 
the constraints (\ref{1-1}), (\ref{1-2}) in this new form is that 
they can be explicitly solved by going to an appropriate 
G-analytic basis in superspace: 
\begin{eqnarray}
  &&x^{\alpha\dot\alpha}_A =  x^{\alpha\dot\alpha} 
+ i(\theta^\alpha_1 \bar\theta^{1\dot\alpha} + \ldots + 
\theta^\alpha_k \bar\theta^{k\dot\alpha} \nonumber \\
&& - \theta^\alpha_{k+1} 
\bar\theta^{k+1\dot\alpha}-\ldots - \theta^\alpha_N 
\bar\theta^{N\dot\alpha} )\;,\nonumber\\ 
  &&\theta^\alpha_I = \theta^\alpha_i u^i_I\;, \quad 
\bar\theta^{\dot\alpha I} = \bar\theta^{\dot\alpha i} 
u_i^I\;.\label{5'} 
\end{eqnarray}
In this basis $W^{12\ldots k}$ becomes an unconstrained function 
of $k$ $\bar\theta$'s and $N-k$ $\theta$'s: 
\begin{equation}\label{5}
   W^{12\ldots k} =  W^{12\ldots k}(x_A,\theta_{k+1},\ldots,\theta_{N},
\bar\theta^1,\ldots,\bar\theta^k, u)\;.
\end{equation}
Altogether it depends on half the number of the odd variables of 
$N$-extended superspace and for this reason we call it Grassmann 
(or G-) analytic. We recall that the notion of Grassmann 
analyticity was first introduced in Ref. \cite{GIO}, still in the 
context of ordinary superspace. In $N=2$ harmonic superspace 
\cite{GIK1} this notion became $SU(2)$ covariant. The 
generalization to $N=3$ was given in Ref. \cite{GIK2} and later on 
to general $N$ in Refs. \cite{hh} (under the name of ``$(N,p,q)$ 
superspace"). 

The massless conformal multiplets describe the ordinary massless 
UIR's of the super Poincar\'e group obtained earlier by the Wigner 
method of induced representations (see, for instance, Ref. 
\cite{str}). The self-conjugate $N=8$ multiplet was obtained by 
the oscillator method in Ref. \cite{gm2}. 

\subsection{Harmonic analyticity as $SU(N)$ \\
irreducibility}

It is important to realize that a G-analytic superfield is an 
$SU(N)$ covariant object only because it depends on the harmonic 
variables. In order to recover the original harmonic-independent 
but constrained superfield $ W^{i_1i_2\ldots 
i_k}(x,\theta,\bar\theta)$ (\ref{1-1}), (\ref{1-2}) we need to 
impose differential conditions involving the harmonic variables. 
The harmonic derivatives are made out of the operators 
\begin{equation}\label{a5}
  \partial^{\,I}_J = u^I_i{\partial\over\partial u^J_i} - 
u^i_J{\partial\over\partial u^i_I} 
\end{equation}
which respect the defining relations (\ref{a2}), (\ref{a2'}). 
These derivatives act on the harmonics as follows: 
\begin{equation}
\partial^I_J u^K_i=\delta^K_J u^I_i~,\qquad
\partial^I_J u^i_K=-\delta^I_K u^i_J~.\label{F9}
\end{equation} 
The diagonal ones $\partial^{\,I}_I$ count the $U(1)$ charges, 
\begin{equation}\label{chc}
  \partial^{\,I}_I u^I_i =  u^I_i\;, \qquad \partial^{\,I}_I u_I^i =  
-u_I^i\;.
\end{equation}
The relation (\ref{a2''}) implies that the charge operators 
$\partial^{\,I}_I$ are not independent, 
\begin{equation}\label{a6}
  \sum_{I=1}^N \partial^{\,I}_I = 0
\end{equation}
(this reflects the fact that we are considering $SU(N)$ and not 
$U(N)$). 

A basic assumption in our approach to the harmonic coset 
$SU(N)/U(1)^{N-1}$ is that any harmonic function is homogeneous 
under the action of $U(1)^{N-1}$, i.e., it is an eigenfunction of 
the charge operators $\partial^{\,I}_I$, 
\begin{eqnarray}\label{eigch}
   &&\partial^{\,I}_I f^{K_1\ldots K_q}_{L_1\ldots L_r}(u) = 
(\delta^{K_1}_I+\ldots + \delta^{K_q}_I - \delta_{L_1}^I -\ldots 
\nonumber \\
&&- \delta_{L_r}^I) f^{K_1\ldots K_q}_{L_1\ldots L_r}(u) 
\end{eqnarray}
(note that the charges  $K_1\ldots K_q;L_1\ldots L_r$ are not 
necessarily all different). Thus it effectively depends on the 
$(N^2-1)-(N-1)=N(N-1)$ real coordinates of the coset  
$SU(N)/U(1)^{N-1}$. Then the actual harmonic derivatives on the 
coset are the $N(N-1)/2$ complex derivatives $\partial^{\,I}_J $, 
$I<J$ (or their conjugates $\partial^{\,I}_J $, $I>J$). 

The set of $N^2-1$ derivatives  $\partial^{\,I}_J $ (taking into 
account the linear dependence (\ref{a6})) form the algebra of 
$SU(N)$: 
\begin{equation}\label{a8}
 [\partial^{\,I}_J,\partial^{\,K}_L ]=\delta^K_J \partial^{\,I}_L -\delta^I_L 
\partial^{\,K}_J~.
\end{equation}
The Cartan decomposition of this algebra $L^+ + L^0 + L^-$ is 
given by the sets 
\begin{eqnarray}\label{a9}
  L^+ &= &\{\partial^{\,I}_J \;, \ I<J\}\;, 
  \quad
L^0 = \{\partial^{\,I}_I\;, \ \sum_{I=1}^N \partial^{\,I}_I = 0 
\}\;,\nonumber \\
  L^- &=& \{\partial^{\,I}_J \;, \ I>J\}\;. 
\end{eqnarray}
It becomes clear that imposing the harmonic conditions 
\begin{equation}\label{a10}
  \partial^{\,I}_J f^{K_1\ldots K_q}_{L_1\ldots L_r}(u) = 0\;, \quad I<J
\end{equation}
on a harmonic function with a given set of charges $K_1\ldots 
K_q;L_1\ldots L_r$ defines the highest weight of an $SU(N)$ irrep. 
In other words, the harmonic expansion of such a function contains 
only one irrep which is determined by the combination of charges 
$K_1\ldots K_q;L_1\ldots L_r$. In fact, not all of the derivatives 
$\partial^{\,I}_J \;, \ I<J$ are independent, as follows from the 
algebra (\ref{a8}). The independent set consists of the $N-1$  derivatives 
\begin{equation}\label{rankder}
  \partial^{\,1}_2\;,\ \partial^{\,2}_3\;,\ \ldots,\ \partial^{\,N-1}_N 
\end{equation}
corresponding to the simple roots of $SU(N)$.
Then the $SU(N)$ defining constraint (\ref{a10}) is equivalent to
\begin{equation}\label{a100}
(\partial^{\,1}_2\;,\partial^{\,2}_3\;,\ldots,\ 
\partial^{\,N-1}_N) f^{K_1\ldots K_q}_{L_1\ldots L_r}(u) = 0\;. 
\end{equation}

The coset $SU(N)/U(1)^{N-1}$ can be parametrized by $N(N-1)/2$ 
complex coordinates. In this case the constraints (\ref{a10}) take 
the form of covariant (in the sense of Cartan) Cauchy-Riemann 
analyticity conditions. For this reason we call the set of 
constraints (\ref{a10}) (or the equivalent set (\ref{a100})) 
harmonic (H-)analyticity conditions. The above argument shows that 
H-analyticity is equivalent to defining a highest weight of 
$SU(N)$, i.e. it is the $SU(N)$ irreducibility condition on the 
harmonic functions. 

As an example, take $N=2$ and the function $f^1(u)$ subject to the 
constraint 
\begin{equation}\label{N2ex}
  \partial^{\,1}_2f^1(u) = 0 \ \Rightarrow \ f^1(u) = f^i u^1_i\;.
\end{equation}
So, the harmonic function is reduced to a doublet of $SU(2)$. 
Similarly, for $N=4$ the function $f^{12}(u)$ is reduced to the 
$\underline 6$ of $SU(4)$. Indeed, the constraints 
$\partial^{\,2}_3 f^{12}(u) = \partial^{\,3}_4 f^{12}(u) = 0$ 
ensure that $f^{12}(u)$ depends on $u^1,u^2$ only, $f^{12}(u)= 
f^{ij}u^1_iu^2_j$. Then the constraint $\partial^{\,1}_2 f^{12}(u) 
=f^{ij}u^1_iu^1_j$ = 0 implies $f^{ij} = -f^{ji}$. 

In the G-analytic basis (\ref{5'}) the harmonic derivatives become 
covariant $D^{\,I}_J$. In particular, the derivatives 
\begin{eqnarray}\label{sptn}
  D^{\,I}_{J}&=& \partial^{\,I}_{J} - i\theta^\alpha_{J} 
\bar\theta^{I\;\dot\alpha}\partial_{\alpha\dot\alpha} - 
\theta_J\partial^I + \bar\theta^I\bar\partial_J\;, \nonumber \\ 
I&=&1,\ldots,k, \ J=k+1,\ldots,N 
\end{eqnarray}
acquire space-time derivative terms. The $SU(N)$ commutation 
relations among the $D^{\,I}_{J}$ are not affected by the change 
of basis. The same is true for the commutation relations of the 
$D^{\,I}_{J}$  with the spinor derivatives: 
\begin{equation}\label{3}
  [D^{\,I}_J,D^K_\alpha]=\delta^K_J D^I_\alpha\;, 
\qquad [D^{\,I}_J,\bar D_{\dot\alpha\; K}]=-\delta^I_K \bar 
D_{\dot\alpha\; J}\;. 
\end{equation}
Using these relations one can see that the H-analyticity 
conditions 
\begin{equation}
  \label{a10'}
  D^{\,I}_J W^{12\ldots k} = 0\;, \quad I<J
\end{equation}
or the equivalent set
\begin{equation}\label{a111}
  (D^{\,1}_2\;,D^{\,2}_3\;,\ldots, D^{\,N-1}_N) W^{12\ldots k} = 0 
\end{equation}
are compatible with the G-analyticity ones (\ref{4}). 

\subsection{Analyticity and massless multiplets: \\ ``Singletons"} 

The constraints of H-analyticity (\ref{a10'}) combined with those 
of G-analyticity  (\ref{4}) have important implications for the 
components of the superfield. First of all, they make each 
component an irrep of $SU(N)$. Take, for example, the first 
component 
\begin{equation}\label{6}
   \phi^{12\ldots k}(x,u) =  W^{12\ldots k}\vert_0
\end{equation}
where $\vert_0$ means $\theta=\bar\theta=0$. The constraints 
$\partial^{\,I}_{I+1}\phi^{12\ldots k}(x,u)=0$, $I=k,\ldots,N$ 
imply that $\phi^{12\ldots k}(x,u)$ takes the form
$$
\phi^{12\ldots k}(x,u) = \phi^{i_1i_2\ldots i_k}(x) u^1_{i_1} 
u^2_{i_2}\ldots u^k_{i_k}\;. 
$$
This is a rank $k$ tensor without any symmetry, i.e. a reducible 
representation of $SU(N)$. Further, the constraint, e.g.,  
$$
\partial^1_2 \phi^{123\ldots k}(x,u) = \phi^{113\ldots k}(x,u) = 
$$
$$
\phi^{i_1i_2i_3\ldots i_k}u^1_{i_1} u^1_{i_2}u^3_{i_3}\ldots 
u^n_{i_k} =0 
$$
removes the symmetric part in the first two indices. Similarly, 
the remaining constraints  (\ref{a10'}) remove all the 
symmetrizations and we find the totally antisymmetric rank $k$ 
irrep of $SU(N)$. 

Another example are the spinor components
\begin{eqnarray}\label{spinco}
&&\chi_{\alpha}^{12\ldots k\; k+1}(x,u) = 
D^{k+1}_{\alpha}W^{12\ldots k}\vert_0 \;, \nonumber \\
&&\bar\psi_{\dot\alpha}^{23\ldots k}(x,u) = \bar 
D_{1\dot\alpha}W^{12\ldots k}\vert_0\;. 
\end{eqnarray}
The same harmonic argument shows that these are harmonic 
projections of the totally antisymmetric components 
$\chi_{\alpha}^{[i_1i_2\ldots i_{k+1}]}(x)$ and 
$\bar\psi_{\dot\alpha}^{[i_2i_3\ldots i_k]}(x)$. 

Further important constraints occur at the level of 2 or more 
$\theta$'s:
\begin{eqnarray}
  D^{I\alpha}D^{J}_\alpha W^{12\ldots k}&=& 0,  
I,J=k+1,\ldots,N,\nonumber \\
 \label{lin2}\\ 
  \bar D_{I\dot\alpha}\bar D_{J}^{\dot\alpha} W^{12\ldots k}&=& 0, 
I,J=1,\ldots,k. \label{lin3} 
\end{eqnarray}
The easiest way to see this is to hit the defining constraint 
(\ref{1-1}) with $D^{k\alpha}$ and then project with harmonics. 

The constraints (\ref{lin2}), (\ref{lin3}) imply that the 
components of the type 
\begin{eqnarray}
  &&\chi^{1\ldots k+p}_{(\alpha_1\ldots \alpha_p)} = D^{k+1}_{\alpha_1}\ldots
D^{k+p}_{\alpha_p} W^{12\ldots k}\vert_0\;, 
 p\leq N-k \nonumber \\
\label{top1}\\ 
  &&\bar\psi^{p+1\ldots k}_{(\dot\alpha_1\ldots\dot\alpha_p)} = 
\bar D_{1\dot\alpha_1} \ldots \bar D_{p\dot\alpha_p} W^{12\ldots 
k}\vert_0\;, \quad p\leq k 
\nonumber \\
\label{top2} 
\end{eqnarray}
are totally symmetric in their spinor indices, i.e. they carry
spin $(p/2,0)$ or $(0,p/2)$, correspondingly. Among them one finds 
the 
\begin{eqnarray}\label{top}
&&\mbox{top spin $({N\over 2}-{k\over 2},0)$:}\quad 
\chi_{(\alpha_1\ldots \alpha_{N-k})} = 
\nonumber \\
&&D^{k+1}_{\alpha_1}\ldots 
D^{N}_{\alpha_{N-k}} W^{12\ldots k}\vert_0 
\end{eqnarray}
which is also an $SU(N)$ singlet. Note that in the case $N=2n$,  
$k=n$ the top spin occurs both as $(n/2,0)$ and $(0,n/2)$ (we call 
this a ``self-conjugate" multiplet). Moreover, if $N=4n$ and 
$k=2n$ one can impose a reality condition on the superfield 
$W^{12\ldots 2n}$ which implies, in particular, that 
\begin{equation}\label{real}
  \chi_{(\alpha_1\ldots \alpha_{2n})} = (\psi_{(\dot\alpha_1\ldots \dot\alpha_{2n})})^* 
\;.
\end{equation}

Next, one can show that all the components of the type 
(\ref{top1}), (\ref{top2}) satisfy massless field equations. 
Indeed, from the constraint (\ref{lin2}) and from G-analyticity it 
follows that 
\begin{eqnarray}
  0&=& \bar D_{k+1\;\dot\beta}D^{k+1\;\alpha_1} D^{k+1}_{\alpha_1}\ldots
D^{k+p}_{\alpha_p} W^{12\ldots k}\nonumber\\ 
  &=&2i\partial_{\dot\beta}^{\alpha_1}D^{k+1}_{\alpha_1}\ldots
D^{k+p}_{\alpha_p} W^{12\ldots k}\nonumber\\ &\Rightarrow& 
\partial_{\dot\beta}^{\alpha_1}\chi^{1\ldots k+p}_{(\alpha_1\ldots \alpha_p)} = 0
 \label{nooo} 
\end{eqnarray}
and similarly for $\bar\psi^{p+1\ldots 
k}_{(\dot\alpha_1\ldots\dot\alpha_p)}$. The leading scalar 
component (\ref{6}) satisfies the d'Alembert equation: 
\begin{eqnarray}\label{dale}
 0 &=& (D^1)^2(\bar D_1)^2 W^{12\ldots k} = 4\Box W^{12\ldots k} \ \nonumber \\
 &\Rightarrow& \ 
\Box\phi^{12\ldots k} = 0.
\end{eqnarray}

Finally, all the components of mixed type, 
\begin{eqnarray}\label{mity}
&&f^{p+1\ldots k+q}_{\dot\alpha_1\ldots\dot\alpha_p 
\alpha_1\ldots\alpha_q}  \nonumber \\
&& = \bar D_{1\dot\alpha_1}\ldots \bar 
D_{p\dot\alpha_p} D^{k+1}_{\alpha_1}\ldots 
D^{k+q}_{\alpha_q}W^{12\ldots k}\vert_0\;, \nonumber \\
&& p\leq k\;, \ q\leq N-k 
\end{eqnarray}
are expressed in terms of the space-time derivatives of lower 
components. Indeed,
\begin{eqnarray}
&& D^1_{k+q} f^{p+1\ldots k+q}_{\dot\alpha_1\ldots\dot\alpha_p 
\alpha_1\ldots\alpha_q}  \nonumber \\
&&= - \bar D_{k+q\dot\alpha_1}\bar 
D_{2\dot\alpha_2}\ldots D^{k+q}_{\alpha_q}W^{12\ldots 
k}\vert_0\nonumber\\ 
  &&= (-1)^{p+q-1}i\partial_{\dot\alpha_1\alpha_q} \bar D_{2\dot\alpha_2}
\ldots D^{k+q-1}_{\alpha_{q-1}}W^{12\ldots k}\vert_0 
 \nonumber \\
&&\Rightarrow \ f^{p+1\ldots k+q-1\; 1
}_{\dot\alpha_1\ldots\dot\alpha_p \alpha_1\ldots\alpha_q} \nonumber \\
&&=(-1)^{p+q-1}i\partial_{\dot\alpha_1\alpha_q}  g^{1\; p+1\ldots 
k+q-1}_{\dot\alpha_2\ldots\dot\alpha_p 
\alpha_1\ldots\alpha_{q-1}} 
\label{nolabel} 
\end{eqnarray}
 
To summarize, the superfield $W^{12\ldots k}$ subject to the 
constraints of G- and H-analyticity has the following component 
content (the derivative terms are not shown):
\begin{eqnarray}
 && W^{12\ldots k} =\phi^{12\ldots k} \nonumber\\
 &&+\bar\theta^1_{\dot\alpha}\bar\psi^{\dot\alpha\; 23\ldots k} + 
\ldots + \bar\theta^k_{\dot\alpha}\bar\psi^{\dot\alpha\; 12\ldots 
k-1} \nonumber\\
 &&+ \theta^\alpha_{k+1}\chi_{\alpha}^{1\ldots k\; k+1}  + \ldots + 
\theta^\alpha_{N}\chi_{\alpha}^{1\ldots k\; N} \nonumber\\ 
 &&+\bar\theta^1_{\dot\alpha} \bar\theta^2_{\dot\beta} 
\bar\psi^{(\dot\alpha\dot\beta)\; 3\ldots k} + \ldots + 
\bar\theta^{k-1}_{\dot\alpha} \bar\theta^k_{\dot\beta} 
\bar\psi^{(\dot\alpha\dot\beta)\; 1\ldots k-2}\nonumber\\ 
 && + \theta^\alpha_{k+1} \theta^\beta_{ k+2} 
\chi_{(\alpha\beta)}^{1\ldots k\; k+1\; k+2} 
 + \nonumber \\
 && \ldots + \theta^\alpha_{N-1} \theta^\beta_{N} 
\chi_{(\alpha\beta)}^{1\ldots k\; N-1\; N}
\ldots\nonumber \\ 
 && + \bar\theta^1_{\dot\alpha_1}\ldots \bar\theta^k_{\dot\alpha_k} 
\bar\psi^{(\dot\alpha_1\ldots\dot\alpha_k)} \nonumber \\
&& +\theta^{\alpha_1}_{k+1} \ldots \theta^{\alpha_{N-k}}_{N} 
\chi_{(\alpha_1\ldots\alpha_{N-k})}  \label{expan} 
\end{eqnarray}
where all the fields belong to totally antisymmetric irreps of 
$SU(N)$ and satisfy the massless field equations
\begin{eqnarray}
  &&\Box \phi^{[i_1\ldots i_k]}=0\;, \nonumber\\
  &&\partial^{\beta\dot\alpha_1} 
\bar\psi_{(\dot\alpha_1\ldots \dot\alpha_p)}^{[i_1\ldots i_{k-p}]} 
= 0\;,\quad 1\leq p \leq k \label{onshell}\\ 
  &&\partial^{\alpha_1\dot\beta}\chi_{(\alpha_1\ldots \alpha_p)}^{[i_1\ldots i_{p}]}
= 0\;, \quad 1\leq p \leq N-k   \nonumber 
\end{eqnarray}
This is the content of an $N$-extended superconformal multiplet of 
the C) series of section 2. It is characterized by the $SU(N)$ 
irrep of the first component (described by the Young tableau 
$m_1=\ldots=m_k=1,\ m_{k+1}=\ldots=m_{N-1}=0$), by its $R$ charge 
\begin{equation}\label{rch}
  r_k={2k\over N}-1
\end{equation}
and conformal dimension $\ell=1$ and by the top spin 
$J_{\scriptsize\rm top} =({N\over 2}-{k\over 2},0)$. 
 
\subsection{Chiral superfields}

The G-analytic superfields considered above contain at least one 
$\bar\theta$. The case of ``extreme" G-analyticity will be the 
absence of any $\bar\theta$'s. These are the well-known chiral 
superfields \cite{fwz} satisfying the constraint
\begin{equation}\label{chir}
  \bar D_{i\;\dot\alpha}W=0 \quad\Rightarrow \quad W=W(x^{\alpha\dot\alpha}_L,
\theta^\alpha_i)
\end{equation}
where
\begin{equation}\label{chb}
  x^{\alpha\dot\alpha}_L = x^{\alpha\dot\alpha} 
- i\theta^\alpha_i\bar\theta^{i\;\dot\alpha}\;. 
\end{equation}
Note that in this case we do not need harmonic variables, since 
G-analyticity involves a subset of odd coordinates forming an 
entire irrep of $SU(N)$, and not a set of $U(1)$ projections. 
Consequently, in order to put such a superfield on shell, we 
cannot use H-analyticity but need to impose a new type of 
constraint: 
\begin{equation}\label{chshell}
  D^{\alpha\; i}D^{j}_\alpha W = 0\;.
\end{equation}
The resulting components are multispinors of the same chirality 
(cf. eq. (\ref{expan})):
\begin{eqnarray}\label{chexpan}
  W &=&\phi + \theta^\alpha_{i}\chi_{\alpha}^{i}  \nonumber \\
   & + & \ldots + 
\theta^{\alpha_1}_{i_1} \ldots \theta^{\alpha_{n}}_{i_n} 
\chi_{(\alpha_1\ldots\alpha_{n})}^{[i_1\ldots i_n]} + \ldots + 
(\theta)^{2N} \chi \nonumber \\
\end{eqnarray}
satisfying  massless field equations. The tops spin is $({N\over 
2} ,0)$. 

The chiral superfields above are scalar, but there exist 
conformally covariant chiral superfields with an arbitrary 
$(J_L,0)$ index of the highest weight: $W_{\alpha_1\ldots 
\alpha_{2J_L}}$. In this case the masslessness condition is 
\cite{HST} $D^{\alpha_1i}W_{\alpha_1\ldots \alpha_{2J_L}}=0$.

\section{Short superconformal multiplets:  bulk `` massless" and ``massive"
states}

In this section we shall concentrate on the case $N=2n$ for 
reasons of simplicity. The analytic superfield $W^{12\ldots 
n}(\theta_{n+1},\ldots,\theta_{2n}, 
\bar\theta^1,\ldots,\bar\theta^n)$ des\-cribes a superconformal 
multiplet characterized by the Young tableau 
$m_1=\ldots=m_n=1,$ $ m_{n+1}=\ldots=m_{2n-1}=0$ of its first component (a Lorentz 
scalar), by its dimension $\ell=1$ and $R$ charge $r=0$ (see 
(\ref{rch})). Now we shall use this multiplet as a building block 
for constructing other ``short" superconformal multiplets.  

The building block $W^{12\ldots n}$ can be equivalently rewritten 
by choosing different harmonic projections of its $SU(N)$ indices 
and, consequently, different sets of G-analyticity constraints. 
This amounts to superfields of the type  
\begin{equation}\label{15}
  W^{I_1I_2\ldots I_n}
(\theta_{J_1},\ldots, \theta_{J_n}, 
\bar\theta^{I_1},\ldots,\bar\theta^{I_n}) 
\end{equation}
where $I_1,\dots,I_n$ and $J_1,\dots,J_n$ are two complementary 
sets of $n$ indices. Each of these superfields depends on $2N=4n$ 
Grassmann variables, i.e. half of the total number of $4N=8n$. 
This is the minimal size of a G-analytic superspace, so we can say 
that the $W$'s are the ``shortest" superfields (superconformal 
multiplets). Another characteristic of these $W$'s is the absence 
of $R$ charges. 

The idea now is to start multiplying different species of the 
$W$'s of the type (\ref{15}) in order to obtain composite objects 
depending on various numbers of odd variables. The sets 
$I_1,\dots,I_n$ can be chosen in $(2n)!/(n!)^2$ different ways. 
However, we do not need consider all of them. The following choice 
of $W$'s and of the order of multiplication covers all possible 
intermediate types of G-analyticity: 
$$
A(p_1,p_2,\ldots,p_{2n-1})= 
$$ 
\begin{eqnarray}
 &&[W^{1\ldots n}(\theta_{n+1 \ldots  2n} 
\bar\theta^{1  \ldots  n})]^{p_1 + \ldots +p_{2n-1}}\nonumber\\ 
  &&\times [W^{1\ldots n-1\; 
n+1}(\theta_{\underline{n}\;{n+2} \ldots  2n} \nonumber \\
&&~~~\times \bar\theta^{1\ldots 
n-1\; \underline{n+1}}) ]^{p_2 + \ldots +p_{2n-1}}\nonumber\\ 
  &&\times [W^{1\ldots n-1\; 
n+2}(\theta_{n\;  {n+1}\; {n+3} \ldots 2n} \nonumber \\
&&~~~\times \bar\theta^{1\ldots 
n-1\; \underline{n+2}} ) ]^{p_3 + \ldots +p_{2n-1}}\nonumber\\ 
  && \cdots     \nonumber\\ 
  &&\times [W^{1\ldots n-1\; 
2n-1}(\theta_{n \ldots {2n-2} \; 2n}\nonumber \\
&&~~~\times \bar\theta^{1\ldots n-1\; 
\underline{2n-1}})]^{p_n + \ldots +p_{2n-1}}\nonumber\\ 
  &&\times [W^{1\ldots n-2\; n\; n+1}(\theta_{\underline{n-1}\; {n+2} 
\ldots  2n}\nonumber \\
&&~~~\times \bar\theta^{1\ldots n-2\; n\; n+1})]^{p_{n+1} + \ldots 
+p_{2n-1}}\nonumber\\ 
  &&\times [W^{1\ldots n-3\; n-1\; n\; n+1}(\theta_{\underline{n-2}\; n+2
 \ldots 2n} \nonumber \\
&&~~~ \times \bar\theta^{1\ldots n-3\; n-1\; n\; n+1}) ]^{p_{n+2} + \ldots 
+p_{2n-1}}\nonumber\\ 
  && \cdots     \nonumber\\ 
  &&\times [W^{13\ldots n+1}(\theta_{\underline{2}\; n+2 
 \ldots 2n} 
\bar\theta^{13\ldots n+1}) ]^{p_{2n-2} +p_{2n-1}}\nonumber\\  
  &&\times [W^{23\ldots n+1}(\theta_{\underline{1}\; {n+2} \ldots  2n} 
\bar\theta^{23\ldots n+1}) ]^{p_{2n-1}} \;.         
\label{verylong} 
\end{eqnarray}

The power $\sum^{2n-1}_{r=k} p_r$ of the $k$-th $W$ is chosen in 
such a way that each new $p_r$ corresponds to bringing in a new 
type of $W$. As a result, at each step a new $\theta$ or 
$\bar\theta$ appears (they are underlined in (\ref{verylong})), 
thus adding new odd dimensions to the G-analytic superspace. The 
only exception of this rule is the second step at which both a new 
$\theta$ and a new $\bar\theta$ appear. So, the series 
(\ref{verylong}) covers all possible subspaces with $4n, 
4n+4,4n+6,\ldots, 8n-2$ odd coordinates (notice once again the 
missing subspace with $4n+2$ odd coordinates). In this sense we 
can say that the G-analytic superfield 
$A(p_1,p_2,\ldots,p_{2n-1})$ realizes a ``short" superconformal 
multiplet.  

The superfield $A(p_1,p_2,\ldots,p_{2n-1})$ should be submitted to 
the same H-analyticity constraints as one would impose on 
$W^{1\ldots n}$ alone,
\begin{eqnarray}\label{alone}
 && D^{\, I}_{I+1}A(p_1,p_2,\ldots,p_{2n-1})=0\;, \nonumber \\
&&I=1,2,\ldots, 2n-1\;.
\end{eqnarray}
This is clearly compatible with G-analyticity since the conditions 
on a generic $A(p_1,p_2,\ldots,p_{2n-1})$ form a subset of these 
on $W^{1\ldots n}$. As before, H-analyticity makes 
$A(p_1,p_2,\ldots,p_{2n-1})$ irreducible under $SU(N)$. Here is 
the structure of Young tableau which corresponds to the first 
(scalar) component of this superfield (and characterizes the 
supermultiplet as a whole): 
\begin{figure}[h]
\setlength{\unitlength}{2mm}
 \begin{center}
\label{doublerep}     
 %\caption{$(m,\ell -m,m)$ representation:}
\begin{picture}(23,8)
%\caption{$(m,\ell -m,m)$}
\put(0,0){\framebox (4,4){\tiny $1$}}
\put(4,0){\framebox (20,4){\tiny $\cdots$}}
\put(24,0){\framebox (4,4){{\tiny $1$}}}
\put(29,1.5){$m_1$}
\put(0,-4){\framebox (4,4){\tiny $2$}}
\put(4,-4){\framebox (16,4){\tiny $\cdots$}}
\put(20,-4){\framebox (4,4){\tiny $2$}}
\put(25,-2.5){$m_2$}
\put(10,-5.5){$\cdots$}
\put(0,-10){\framebox (4,4){\tiny $k$}}
\put(4,-10){\framebox (12,4){\tiny $\cdots$}}
\put(16,-10){\framebox (4,4){\tiny $k$}}
\put(21,-8.5){$m_k$}
\put(8,-11.5){$\cdots$}
\put(0,-16){\framebox (4,4){\tiny 2n-1}}
\put(4,-16){\framebox (8,4){\tiny $\cdots$}}
\put(12,-16){\framebox (4,4){\tiny 2n-1}}
\put(17,-14.5){$m_{2n-1}$}
\end{picture}
  \end{center}
\end{figure}
\vskip 3cm \noindent The top row is filled with indices projected 
with $u^1_i$ (hence the symmetrization among them), the second row 
- with $u^2_i$, etc. The harmonic conditions (\ref{alone}) remove 
all the symmetrizations among indices belonging to different 
projections (rows). By counting the number of occurrences of the 
projection $1$ in (\ref{verylong}), we easily find the relation 
\begin{equation}\label{m-l}
  m_1 = \ell - p_{2n-1}
\end{equation}
where $\ell$ is the total number of $W$'s (equal to the dimension 
of the superfield $A$, since $\ell_W=1$). Another simple counting 
shows the relation 
\begin{equation}\label{sum-l}
  \sum^{2n-1}_{k=1}m_k = n\ell = {N\over 2}\ell\;.
\end{equation}
If the last $W$ in (\ref{alone}) is not present there is an 
additional relation among the Young tableau labels: 
\begin{equation}\label{simre}
  p_{2n-1}=0 \quad \Rightarrow \quad  m_1  = {2\over 
N}\sum^{2n-1}_{k=1}m_k\;.
\end{equation}
\vfill\eject

Finally, introducing the Dynkin labels $[a_1,\ldots,$ $a_{2n-1}]$ 
where $a_{1}=m_{2n-1}$ and $a_k = m_{2n-k+1}-m_{2n-k}$ for $k\geq 
2$, we find 
\begin{eqnarray}
  && a_{1}= \sum_{k=n}^{2n-1} p_k\;,  \nonumber\\
  && a_{2}= p_{n-1}\;, \quad  \ldots\;, \quad  a_{n-2}= p_{3}\;,  \nonumber\\
  && a_{n-1}= p_2 + \sum_{k=n+1}^{2n-1} (k-n)p_{k}\;, \nonumber\\ 
  && a_{n}= p_{1} \;,  \label{DL}  \\
  &&  a_{n+1}= (n-2)\sum_{k=n+1}^{2n-1} p_k + 
  \sum_{k=2}^{n} (k-1)p_{k}\;,   \nonumber\\
  && a_{n+2}= p_{n+1} \;, \quad  \ldots\;, \quad 
  a_{2n-1}= p_{2n-2}\;.   \nonumber
\end{eqnarray}

\setcounter{equation}0 
\section{Conclusion}

In this paper we studied representations of four-dimensional superconformal
algebras with an arbitrary number of supersymmetries.

This analysis also provides the classification of short multiplets of
superalgebras on AdS$_5$ and in particular ``massless" and ``massive" fields in
anti-de Sitter geometries, in terms of boundary ``composite" operatprs

\section*{Acknowledgements}

This  work  has been supported in part by the 
European Commission TMR programme ERBFMRX-CT96-0045 (Laboratori 
Nazionali di Frascati, INFN) and by DOE grant DE-FG03-91ER40662, 
Task C.

%%%%%%%%%%%%%%%%%%%%%%%%%%%%%%%%%%%%%%%%%%%%%%%%

\end{document}